\documentclass[twocolumn,prl,superscriptaddress,showpacs,preprintnumbers,amsmath,amssymb,amsmath,showpacs,nofootinbib]{revtex4}

\usepackage{graphicx}
\usepackage{epsf}
\usepackage{epsfig}

\usepackage{subfigure}
\usepackage{amsmath,amssymb}
\usepackage{rotating}
\usepackage{bm}
\usepackage{color}

\newcommand{\dd}{{\mathrm{d}}}
\newcommand{\Z}{{\mathbf{Z}}}

\def\di{\displaystyle}

\def\bg{\begin{eqnarray}\begin{array}{rcl}\displaystyle}
\def\eg{\end{array} &\di    &\di   \end{eqnarray}}
\def\bm#1{\begin{eqnarray}\begin{array}{#1}\di}
\def\bmo#1{\begin{eqnarray*}\begin{array}{#1}\di}
\def\bml#1#2{\begin{eqnarray}\begin{array}{#1}\label{#2}\di}
\def\bgo{\begin{eqnarray*}\begin{array}{rcl}\displaystyle}
\def\ego{\end{array} &\di    &\di \nonumber  \end{eqnarray*}}

\def\btensor#1#2{\left#1\begin{array}{#2}\di}
\def\brtensor#1#2#3{\ren#3\left#1\begin{array}{#2}}
\def\botensor#1#2{\renew\left#1\begin{array}{#2}}
\def\etensor#1{\end{array}\right#1}

\def\eq#1{(\ref{#1})}
\def\Eq#1{Eq.~(\ref{#1})}

\def\tr{{\rm tr}}
\def\Tr{{\rm Tr}}

\def\id{1\!\mbox{l}}

\def\s0#1#2{\mbox{\small{$ \frac{#1}{#2} $}}}
\def\0#1#2{\frac{#1}{#2}}

\def\CB{{\mathcal B}}
\def\CC{{\mathcal C}}
\def\CD{{\mathcal D}}

\def\CP{{\mathcal P}}

\definecolor{yuji}{rgb}{0,0.4,0.7}
\definecolor{jan}{rgb}{0.7,0,0.4}

\newcommand{\nn}{\nonumber}

\begin{document}

\title{Majorana fermions and CP-invariance of chiral
  gauge theories on the lattice}

\pacs{11.15.-q,11.15.Ha,11.30.Er,11.30.Rd\vspace{-.2cm}}

\author{Yuji Igarashi } 
\affiliation{Faculty\ of\ Education, 
 Niigata\ University, Ikarashi, 
{950-2181}, Niigata, Japan}
\author{Jan M.~Pawlowski}
\affiliation{Institut f\"ur Theoretische Physik, Universit\"at
  Heidelberg, D-69120 Heidelberg, Germany}

\begin{abstract}
  The construction of massless Majorana fermions with chiral Yukawa
  couplings on the lattice is considered. We find topological
  obstructions tightly linked to those underlying the Nielsen-Ninomiya
  no-go theorem. In contradistinction to chiral fermions the
  obstructions originate only from the combination of the Dirac action
  and the Yukawa term. These findings are used to construct a chirally
  invariant lattice action. We also show that the path integral of
  this theory is given by the Pfaffian of the corresponding Dirac
  operator. As an application of the approach set-up here we construct
  a CP-invariant lattice action of a chiral gauge theory, based on a
  lattice adaptation of charge conjugation and parity transformation in
  the continuum.
\end{abstract}

\maketitle

\noindent {\bf  Introduction.---} 
Massive neutrinos can be incorporated
in the Standard Model with Majorana fermions that become massive via
spontaneous symmetry breaking. This mass generation necessitates a
chirally symmetric Yukawa term. A lattice approach to the related
physics problems has to be based on an appropriate lattice formulation
of Majorana fermions in the presence of chiral symmetry
\cite{Ginsparg:1981bj,Neuberger:1997fp,Hasenfratz:1998ri,Luscher:1998pq},
see also \cite{Fujikawa:2002fi,Suzuki:2004ht,Pawlowski:2007ua}.
Majorana fermions with chiral symmetry also play a key r$\hat{\rm
  o}$le for physics beyond the Standard Model, in particular in
supersymmetric theories. Realisations of lattice supersymmetry face
additional problems since space-time lattices are generally
irreconcilable with supersymmetry, see e.g.\
\cite{Dondi:1976tx,Curci:1986sm,Fujikawa:2002ic,
  Sugino:2003yb,D'Adda:2007ee,Bergner:2008ws,Kato:2008sp}, for reviews
see \cite{Feo:2002yi,Kaplan:2003uh,Giedt:2007hz}.

In \cite{Pawlowski:2007ua} an approach to chirally
coupled Majorana fermions was initiated. Here we extend this work,
evaluate the related obstructions and provide a construction of
chirally coupled Majorana fermions. This construction may also prove
useful for the construction of supersymmetric theories on the
lattice. Within a lattice formulation, chiral symmetry becomes
non-trivial due to the Nielsen-Ninomiya no-go theorem
\cite{Karsten:1980wd,Nielsen:1980rz,Karsten:1981gd,Friedan:nk,%
  Jahn:2002kg}, and we expect related obstructions for chirally
coupled Majorana fermions.  Indeed there appears a certain conflict
between the definition of the Majorana fermions and lattice chiral
symmetry in the presence of Yukawa couplings.  The conflict is closely
related to the requirements of locality and that of avoiding species
doubling, which are the basic issues of lattice chiral symmetry. It
also causes an obstruction in constructing the simplest supersymmetric
model, the Wess-Zumino model on a lattice, and in showing CP
invariance of chiral gauge theory, see
e.g.\cite{Luscher:1998du,Fujikawa:2002vj,Hasenfratz:2005ch}.

In the first section we recapitulate the continuum formulation of
Majorana fermions. Then we discuss the specialities of a lattice
formulation, in particular the necessary doubling of degrees of
freedom, and the topological obstructions inherited from the demand of
chiral invariance. The findings are used to construct Majorana
fermions including the proof of the Pfaffian nature of the lattice
path integral. Finally we construct a CP-invariant lattice action of a
chiral gauge theory based on a lattice extension of charge conjugation
and parity transformation.\\[-1ex]

\noindent {\bf Majorana fermions.---} 
Majorana fermions are neutral fermions and hence obey a reality
constraint.  In four-dimensional Euclidean space-time the charge
conjugation operator $C$ has the properties
\begin{eqnarray}\nonumber 
C \gamma_\mu C^{-1}=-\gamma_\mu^T\,, &\qquad &
C \gamma_5 C^{-1}=\gamma_5^T\,, \\[1ex] C^\dagger C=\id 
\, \,, &\qquad& C^T= -C\,, 
\label{eq:charge}  \end{eqnarray} 
and the reality constraint for Majorana fermions reads  
\begin{eqnarray}\label{eq:reality} 
\psi^*=B\psi\,,
\end{eqnarray}
where {$C=\gamma_5\,B$}. However, \eq{eq:charge} implies that $B^*
B=-\id$ and hence we cannot implement the reality constraint
\eq{eq:reality} as it fails to satisfy the consistency condition
$\psi^{**}=\psi$. This problem is circumvented by doubling the degrees
of freedom which suffices to implement the reality constraint with
\begin{eqnarray}\label{eq:symplecticreality} 
\psi^*= \CB \psi \,, \qquad {\rm with} \qquad \psi=\btensor{(}{c} \psi_1 \\ 
\psi_2 \etensor{)}\,,  \end{eqnarray} 
and 
\begin{eqnarray*} 
\CB= \btensor{(}{ccc} 0 & & B \\ -B & &  0 \etensor{)} 
\,.
\end{eqnarray*} 
The symplectic structure of $\CB$ leads to $\CB^*\CB=\id$ with $B^*
B=-\id$. Thus the reality constraint, $\psi^{**}=\psi$, is satisfied.
The corresponding charge conjugation operator is provided by
\begin{eqnarray}\label{eq:Ccharge} 
\CC= \btensor{(}{ccc} 0 & & C \\ C & &  0 \etensor{)} = {\Gamma_5\,\CB},\quad 
{\rm with}\quad \Gamma_5=\btensor{(}{ccc} {\gamma_5} & & 0 \\
 0 & &   {-\gamma_5} 
\etensor{)}\,.
\end{eqnarray} 
The above properties of the symplectic Majorana fermion $\psi$ fix its
behaviour under chiral rotations,
\begin{eqnarray}\label{eq:chiral} 
\psi\to (1+ i \epsilon\Gamma_5)\psi\,,  
\end{eqnarray} 
which introduces a relative minus sign in the chiral rotation of
$\psi_1$ and $\psi_2$. We are now in the position to construct a
chirally invariant Majorana action.  We summarise the necessary
properties,
\begin{eqnarray}\nonumber 
\CC={\Gamma_5\,\CB},&\qquad& 
\CC \Gamma_5\CC^{-1}=-\Gamma_5^T\,, \\[1ex]
\CC^\dagger \CC=\id 
\, \,, &\qquad& \CC^T= -\CC\,,  
\label{eq:Cprops} \end{eqnarray} 
and construct the corresponding chirally invariant Majorana action  
\begin{eqnarray}\nonumber 
S_{D}[\psi] &=&\int d^4 x\, \psi^T \CC \CD\psi\\ &=& 
\int d^4 x\, \left(\psi_1^T C D\psi_1+\psi_2^T C D
\psi_2\right)
\label{eq:majocont} \end{eqnarray} 
with 
\begin{eqnarray}\label{eq:majocont1}
\CD=
  \btensor{(}{ccc} 0 & & D \\ D & &  0 \etensor{)}\,, \qquad {\rm and} 
\qquad (\CC\CD)^T=-\CC\CD \,. 
\end{eqnarray} 
The action \eq{eq:majocont} could also be obtained by a Majorana
reduction, see e.g. \cite{Nicolai:1978,van Nieuwenhuizen:1996}. Note
that chiral invariance of \eq{eq:majocont} is trivial, because
\begin{eqnarray}\label{eq:majoproperty}
\psi_i^T C \gamma_{5} D\psi_i =0 \qquad (i=1,2)\,.
\end{eqnarray} 
We also remark that skew symmetry of
$\CD$ is not required but only the skew-symmetric part of $\CD$
contributes to the action $S$. The definitions \eq{eq:majocont1} imply
\begin{eqnarray} \label{eq:skew} 
(C\, D)^T=-C\,D \,,\qquad {\rm and} \qquad  D^*={B D B^{-1}\,},
\end{eqnarray} 
for the Dirac operator $D$. The combined properties \eq{eq:skew} hold 
for the standard chiral Dirac operator with
\begin{eqnarray}\label{eq:schiral} 
\gamma_5 D+D\gamma_5 =0\,,
\end{eqnarray} 
such as $D=\gamma_\mu \partial_\mu$, for which \eq{eq:skew} can be
deduced from \eq{eq:charge}. The action \eq{eq:majocont} is chirally 
invariant under a chiral transformation with \eq{eq:chiral} if 
\begin{eqnarray}\label{eq:chiraldirac} 
\Gamma_5 \CD-\CD\Gamma_5 = 0\,, 
\end{eqnarray} 
which is valid for $\CD$ with the standard Dirac operator satisfying
\eq{eq:schiral}.  Finally we remark that the action \eq{eq:majocont}
is real, as follows from \eq{eq:skew}.  It is instructive to make this
reality explicit by rewriting the action \eq{eq:majocont} with the
help of the above relations,
\begin{eqnarray}\label{eq:majocontreal} 
S_{D}[\psi]
=\int d^4 x\, \left(\psi_2^\dagger \gamma_5 D\psi_1 +\psi_1^\dagger 
\gamma_5 D^\dagger \psi_2\right)\,. 
\end{eqnarray} 
Note that \eq{eq:majocontreal} is even real for unconstrained Dirac
fermions $\psi_1,\psi_2$ in contrast to \eq{eq:majocont}.  

The eqs.~\eq{eq:majocont},\eq{eq:majocontreal} provide the action of a
Euclidean Majorana fermion. For the chirally coupled Yukawa theory we
are also interested in Weyl fermions which we can construct from the
components of the Majorana spinor $(\psi_1,\psi_2)$. To that end let
us introduce chiral projection operators related to $\Gamma_5$ in
\eq{eq:chiral},
\begin{eqnarray}\label{eq:P}
  \CP=\012 (1+\Gamma_5)=\btensor{(}{ccc} P & & 0 \\ 0 & &  
1-P \etensor{)} \,,\end{eqnarray} 
with 
\begin{equation*} {P= \012 (1+ \gamma_5)}\,.
\end{equation*} 
The chiral projection operators $\CP$, $(1-\CP)$ allow us to project
on right-handed and left-handed spinors. 
For a given pair of $\psi_1$ and $\psi_2$, we can construct a Weyl action 
taking its off-diagonal combination:
\begin{eqnarray}\label{eq:Weylcontinuum} 
S_{W}[\psi]&=&\int d^4 x\, \psi^T \CC 
\btensor{(}{ccc} 0 & & 0 \\ 1 & &  0 \etensor{)}
\CD \CP\psi\nn\\
&=&\int d^4 x\, \psi_2^T C  D P\psi_1 ,. 
\end{eqnarray} 
We shall use this type of action to discuss CP invariance of a chiral
gauge theory on the lattice.  

For the construction of a chirally invariant Yukawa term we have to
couple chiral projections of the Majorana fermions to the scalar
field, 
\begin{eqnarray}\nonumber 
S_{Y}[\psi,\phi] &= &g  
\int d^4 x\Bigl(\, {\psi}^{T}\CC \CP 
\phi^{\dagger} ~  {\psi}+ {\psi}^{T}\CC(1-\CP) 
\phi  {\psi}\Bigr)\\
&=&g\int d^4 x \Bigl( 
{\psi_1}^{T}C\, P 
\varphi ~ {\psi_1}+ {\psi_1}^{T}C\, (1-P) 
\varphi^{\dagger}  {\psi_1}\nonumber\\
&&   +{\psi_2}^{T}C\, P 
\varphi^{\dagger} ~ {\psi_2}+ {\psi_2}^{T}C\, (1-P) 
\varphi  {\psi_2}  \Bigr) \,. 
\label{eq:S_Y}\end{eqnarray}
In \eq{eq:S_Y} $\varphi$ is a complex scalar, and $\phi$ is defined by  
\begin{eqnarray}\label{eq:phi}
\phi= \btensor{(}{ccc} 0 & & \varphi \\ \varphi & &  0 \etensor{)}\,, \qquad 
{\rm with} \qquad  \phi \to (1{-}2i\epsilon)\phi\,.
\end{eqnarray} 
Note that the scalar field $\phi$ is off-diagonal and hence does not 
commute with the projection operators, we have e.g.\, $\CP
\phi^\dagger =\phi^\dagger (1-\CP)$. The action
$S_{D}[\psi]+S_Y[\psi,\phi]$ is invariant under the
transformation \eq{eq:chiral} of the fermions and that in \eq{eq:phi}
for the scalar field $\phi$ related to $\varphi\to (1
{-}2i\epsilon)\varphi$. This concludes our brief summary of chirally
coupled Majorana fermions in the continuum. Due to chiral symmetry,
and in particular the use of chiral projections in \eq{eq:S_Y} we
expect obstructions for putting
the above theory on the lattice.\\[-1ex]

\noindent {\bf Topological obstructions on the lattice.---}
Chiral symmetry on the lattice differs from that in the continuum as
consistent chiral transformations necessarily depend on the Dirac
operator.  Hence we first discuss the properties of the lattice
version of the free Dirac action \eq{eq:majocont}
\begin{eqnarray}\label{eq:majolattice}
S_{D}[\psi] =
\sum_{x,y\in\Lambda} \Psi^T(x) \CC \CD(x-y) \Psi(y)\,, 
\end{eqnarray} 
with the lattice Dirac operator $D(x-y)$ used in the definition of
$\CD$ as defined in \eq{eq:majocont1}, and $\Psi=(\Psi_1,\Psi_2)$
obeys the symplectic relatity constraint \eq{eq:symplecticreality}. Assume for
the moment that $D(x-y)$ is of Ginsparg-Wilson type
\cite{Ginsparg:1981bj} with
\begin{eqnarray}\label{eq:GWga} 
\gamma_5 {D}+{D}\gamma_5= a {D}\gamma_5{D}\,,  
\end{eqnarray}
related to a vector-symmetric blocking procedure.  Then the chiral
transformation
\begin{eqnarray}\label{eq:GWchiral} 
{ \Psi\to \left[1+ i \epsilon\Gamma_5 \left(1-\012 a
\btensor{(}{ccc} 0 & & 1 \\ 1 & &  0 \etensor{)}
\CD\right)\right]\Psi}\,, 
\end{eqnarray} 
is an invariance of \eq{eq:majolattice}. However, smooth chiral
projections $P$ and $\CP$ cannot be constructed, which is reflected in
the fact that the transformation \eq{eq:GWchiral} vanishes at the
doublers.  This is a consequence of the well-known Nielsen-Ninomiya
no-go theorem
\cite{Karsten:1980wd,Nielsen:1980rz,Karsten:1981gd,Friedan:nk,Jahn:2002kg},
which provides obstructions for putting chiral fermions on the
lattice.  Ginsparg-Wilson fermions \cite{Ginsparg:1981bj} circumvent
the no-go theorem with a modified chiral symmetry \eq{eq:GWga}, which
can be reformulated as
\begin{eqnarray}\label{eq:GW} 
  \gamma_5 {D}+{D}\hat \gamma_5=0, \qquad {\rm with} 
\qquad \hat \gamma_5=
\gamma_5 (1-a {D })\,, 
\end{eqnarray} 
and chiral projections 
\begin{eqnarray}\label{eq:project} 
{P= \s012 (1 {+} \gamma_5),\qquad \hat P =
\s012(1 {+} \hat \gamma_5)}\,.
\end{eqnarray}
The general case \cite{Jahn:2002kg,Bergner:2008ws}, going beyond
Ginsparg-Wilson fermions, including e.g.\cite{Fujikawa:2000my}, only
resorts to general chiral projections $P,\hat P$, which are
compatible:
 \begin{eqnarray}\label{eq:genchiral} 
{(1-P)\, D = D\, \hat P}\,. 
\end{eqnarray}
It has been shown in \cite{Jahn:2002kg} that projection operators
$P,\hat P$ carry a winding number that is related to the
total chirality $\chi$ of the system at hand,
 \begin{eqnarray}\label{eq:chirality} 
\chi  = n[\hat P]-n[1-P]\,, 
\end{eqnarray} 
with 
\begin{equation*}
n[P] \equiv \frac{1}{{2}!}\left(\frac{i}{2\pi}\right)^{2} 
\int_{T^{4}} \tr\,P (\dd P)^{4} \,.
\in \Z \,, 
\end{equation*}
if $\hat P\Psi =\Psi$ in the action. \Eq{eq:chirality} also entails
that for odd chirality $\chi$, $\hat P\Psi$ and $P \Psi$ live in
topologically different spaces, and hence have to be different. In the
present case the total chirality $\chi$ is even due to the symplectic
construction. The continuum Yukawa action, however, contains
projection operators $\CP,1-\CP$ with $P,1-P$ on chiral sub-spaces
with $\CP \Psi\neq \Psi$, that is on fermionic sub-systems with odd
chirality.  Thus we have to worry about the use of projection
operators in the Yukawa action $S_Y$.

The first question that arises in this context is whether the lattice
Yukawa action can be constructed such that it is left invariant under
the chiral transformations \eq{eq:GWchiral}, and tends toward the
continuum action. This would require the existence of a smooth
operator $\tilde P$ which reduces $\tilde P\to 1-P$ in the continuum
limit, and ensures invariance of the Yukawa term under the combined
transformation \eq{eq:phi} and \eq{eq:GWchiral}. 
Furthermore, there is no
transformation of the scalar field $\phi$ that could absorb a
momentum-dependent transformation of $\tilde P \Psi$.  Hence, a
necessary condition for the invariance of the Yukawa term is a
transformation of $\tilde P \Psi$ that is independent of the Dirac
operator $D$,
\begin{equation}\label{eq:indep}
\tilde P \Psi\to \pm \epsilon M \tilde P \Psi\,,  
\end{equation} 
with constant matrix $M$. As $\gamma_5(1-\0{a}{2}D)$ is not normalised
and even vanishes at the doublers such an operator $\tilde P$ cannot
exist, even if one relaxes the projection property $\tilde P^2=\tilde
P$.

In turn it is required that the chiral transformation must be
compatible with the projection operators used in the Yukawa term. This
already excludes \eq{eq:GWchiral}. Without loss of generality we can
restrict ourselves to the chiral transformation
\begin{eqnarray}\label{eq:chiraltrafo}
\Psi\to (1+i \epsilon \hat\Gamma_5)\Psi\,, 
\quad 
{{\rm with} \quad \hat\Gamma_5=\btensor{(}{ccc} \hat\gamma_5 & & 0 \\ 0 
& &  -\hat\gamma_5 
\etensor{)}}\,,
\end{eqnarray} 
and hence 
\begin{eqnarray*}
\Psi^T\hat\CC 
\to \Psi^T\hat\CC\,[
\hat\CC^{-1} (1+i \epsilon \hat\Gamma_5^T) \hat\CC]\,,
\end{eqnarray*} 
where $\hat\CC$ is a lattice generalisation of $\CC$.  Then, chiral
invariance of the action $S$ in \eq{eq:majolattice} leads to the
constraint
\begin{eqnarray}\label{eq:constraintG} 
\hat\CC^{-1} \hat\Gamma_5^T \hat\CC = -\Gamma_5\,,\qquad {\rm with} \qquad 
\Gamma_5 \CD= \CD\hat\Gamma_5\,. 
\end{eqnarray} 
We conclude that invariance of the lattice action \eq{eq:majolattice}
under the chiral transformations \eq{eq:chiraltrafo} would require
\begin{eqnarray}\label{eq:compatible}
 \hat \gamma_5^T =\hat C \gamma_5 {\hat C}^{-1} \,, 
\end{eqnarray}
which maps $\hat\gamma_5$ carrying the winding number $n[\hat P]$ to
$\gamma_5$ carrying the winding number $n[P]$. Note that using
different $\gamma_5$'s in the definition of $\Gamma_5$ still leads to
the same conclusion \eq{eq:compatible}. In order to elucidate this
obstruction we use Ginsparg-Wilson fermions as an example.  There the
relation \eq{eq:compatible} reads
\begin{eqnarray}\label{eq:compatibleGW} 
(1-a D^T)\gamma_5^T =\hat C \gamma_5 {\hat C}^{-1} \,, 
\end{eqnarray}
with a possible solution 
\begin{eqnarray}\label{eq:solGW} 
\hat C =C (1-\s012 a D)\,.
\end{eqnarray} 
The $\hat C$ in \eq{eq:solGW} has zeros at the doublers, and the
relative winding number is carried by these zeros. Inserting a lattice
$\hat C$ in \eq{eq:solGW} into the action \eq{eq:majolattice} we
encounter zeros or singularities of the operator ${\hat C}^{-1} D$ at
the positions of the doublers. This brings back the doubling problem.
Consequently, we have to use independent Majorana fields $\psi$,
$\Psi$ with different chiral transformation properties for the
construction of Majorana actions. At the same time, our discussion
given here also suggests the possibility of defining a lattice
generalisation of charge conjugation as a discrete transformation
between these topologically different spinors. We shall discuss first
the construction of Majorana actions, and then how to define charge
conjugation in connection with CP invariance of chiral gauge theories.\\[-1ex]

\noindent {\bf Construction of Majorana actions.---} Now we are in the
position to construct chirally coupled Majorana fermions on the
lattice.  In line with the arguments of the last section we introduce
two sets of symplectic Majorana fermion, $\psi$ and $\Psi$.  
Then
chiral invariance is easily arranged for with appropriate, different,
chiral transformations for $\psi$ and $\Psi$ respectively.
Furthermore we have to ensure that our path integral results in
Pfaffians of the Dirac operator which signals Majorana fermions. The
corresponding lattice action reads
\begin{eqnarray}\label{eq:majolatticedoub}
S_{D}[\psi,\Psi] =
\sum_{x,y\in\Lambda} {\psi}^T(x) \CC \CD(x-y) \Psi(y)\,,  
\end{eqnarray} 
with the Yukawa term
\begin{widetext}
\begin {eqnarray} \label{eq:S_Ylattice}
S_{Y}[\psi,\Psi,\phi] &= &g  
\sum_{x,y\in\Lambda}\left(\, {\psi}^{T}\CC \CP 
 ~  {\phi}^\dagger (1-\hat \CP) {\Psi}+ {\psi}^{T}\CC (1-\CP) 
   {\phi} \hat\CP {\Psi}\right)
\\
&=& {g\sum_{x,y\in\Lambda} \bigl( 
{\psi_1}^{T}C\, P 
 ~ \varphi \hat P {\Psi_1}+ {\psi_1}^{T}C\, (1-P) 
{\varphi^{*}} (1-\hat P) {\Psi_1}} {+{\psi_2}^{T}C\, P \varphi^*
 ~ \hat P  {\Psi_2}
+ {\psi_2}^{T}C\, (1-P) 
\varphi (1-\hat P)  {\Psi_2}\bigr)\,,}
\nonumber \end{eqnarray} 
\end{widetext}
where 
\begin{equation}
\begin{array}{rclcrcl}
{\cal P}&=&(1 +
\Gamma_{5})/2\,,&\qquad & P &=& 
(1+\gamma_{5})/2\\ [1ex]
\hat{\cal P}&=& (1+ \hat\Gamma_{5})/2\,, &\qquad & 
{\hat P}&=&(1+\hat\gamma_{5})/2\,.
\label{eq:PhatP}\end{array}\end{equation}. 
Note that
the scalar field $\varphi$ does not commute with the projection
operators $P,\hat P$, as they depend in general on the Dirac operator. 
Hence, in \eq{eq:S_Ylattice} the combinations such as $ P 
~ \varphi \hat P$ cannot be reduced to  $P\hat P 
~ \varphi$. We also remark 
that the construction given here differs from that in 
\cite{Luscher:1998pq,Kikukawa:20005}. 
There additional Majorana fermions 
were introduced as static auxiliary fields, and are used to 
construct a chirally invariant Yukawa term, see also
\cite{Elliott:2008jp,Giedt:2007qg,Poppitz:2009gt}.

The action $S_{D}+S_Y$ is invariant under the chiral 
transformations
\begin{eqnarray}\nonumber 
&&\Psi\to (1+i\epsilon\hat\Gamma_5) \Psi\,,\qquad 
\psi\to  (1+i\epsilon\Gamma_5 )\psi\,,\\[1ex] &&\phi\to (1 -
2 i\epsilon) \phi\,.
\label{eq:chirallat} \end{eqnarray} 
The action $S+S_Y$ reduces to the continuum action in the continuum
limit, but with a doubling of the field content. This doubling can be
removed by appropriate prefactors in the action, or by simply taking
roots of the generating functional $Z$. However, it is left to prove
the Pfaffian nature of the path integral. Since we have doubled the
degrees of freedom we could have constructed a Dirac fermion out of
two Majorana fermions. 
For the proof it is sufficient to concentrate  
on the path integral
of the free Majorana action \cite{Suzuki:2000ku,Inagaki:2004ar}
including a mass term for dealing with the zero modes: 
\begin{equation}\label{eq:pathmajo}
Z=\int \prod_x d\psi_1 d\psi^*_1  d\Psi_1 d{\Psi_1}^*\, 
\exp -(S_{D}[\psi,\Psi]+ S_{m}[\psi,\Psi])
\,,  
\end{equation}
where the Majorana action \eq{eq:majolatticedoub} couples $\psi_1$ to 
$\Psi_1$ and $\psi_2$ to 
$\Psi_2$ respectively,  
\begin{eqnarray}\label{eq:actionpi}
  S_{D}[\psi,\Psi]&=& \sum_{x,y\in\Lambda} {\psi}^T(x) 
  \CC \CD(x-y) \Psi(y)
  \nn\\[1ex] \nonumber 
  &=& \sum_{x,y\in\Lambda} \Bigl({\psi}_{1}^{T}(x) C D(x-y) ~
  {\Psi}_{1}(y) \\[1ex] 
  &&+  {\psi}_{2}^{T}(x) C D(x-y) ~
  {\Psi}_{2}(y) \Bigr)\,.    
\end{eqnarray} 
This also applies to the regularising mass term 
\begin{eqnarray}\label{eq:actionpimass}
&&\hspace{-1cm}S_{m}[\psi,\Psi]= 
{i m}\sum_{x\in\Lambda} {\psi}^T(x) \CC 
\btensor{(}{ccc} 0 & & 1 \\ 1 & &  0 \etensor{)} 
(\Gamma_5 \Psi)(x)\,
\\
&=&\hspace{.2cm}
 im \sum_{x} \left({\psi}_{1}^{T}(x) C \gamma_{5} {\Psi}_{1}(x) 
- {\psi}_{2}^{T}(x) C \gamma_{5} {\Psi}_{2}(x) \right)\,.\nonumber 
\end{eqnarray}
The Pfaffian is computed in terms of eigenvalues of the Dirac operator 
$\CC \CD$. Hence we proceed by expanding the fermionic fields $\psi,\Psi$, 
and consequently the action \eq{eq:actionpi},
\eq{eq:actionpimass} 
in terms of eigenfunctions of $\CC \CD$. The operator $\CC \CD$ 
can be constructed from the 
hermitian operator $H= \gamma_{5} D$ and its corresponding eigenfunctions 
${\varphi}_{n}$ by
\begin{equation}\label{eq:eigenvect} 
  \sum_{y} H(x-y) {\varphi}_{n}(y) = \lambda_{n} {\varphi}_{n}(x)\,,
\end{equation} 
with  
\begin{equation*} 
\left( {\varphi}_{n},~{\varphi}_{m}\right) = 
  \sum_{x} {\varphi}_{n}^{\dagger}(x)
  {\varphi}_{m}(x) = \delta_{n,m}\,. 
\end{equation*}
It straightforwardly follows that there is a double degeneracy in 
these equations: from the eigenfunctions $\varphi$ we can construct 
linearly independent eigenfunctions $\phi_n=\gamma_5 C^{-1} \varphi_n^*$
with eigenvalues $\lambda_{n}$, that is 
$H(x) \phi_{n}(x) = \lambda_{n} \phi_{n}(x)$, and 
$(\phi_{n},\varphi_{m})=0$.  Therefore, the
fields $\Psi_{1}$ and $\psi_{1}$ can be expanded as
\begin{eqnarray}\label{eq:expan}
{\Psi}_{1}(x) &=& \sum_{n} \left[ {\varphi}_{n}(x)c_{n}  
+ {\phi}_{n}(x) b_{n} \right] \nn\\
{\psi}_{1}(x) &=& \sum_{n} \left[ {\varphi}_{n}(x)c_{n}^{\prime} + 
  {\phi}_{n}(x) b_{n}^{\prime}  \right]\,.
\end{eqnarray}
Inserting this expansion into the action \eq{eq:actionpi},\eq{eq:actionpimass} 
we are led to 
\begin{eqnarray}\nonumber 
&&\hspace{-1cm}  S_{D}[\psi,\Psi]+ S_{m}[\psi,\Psi] \\[1ex]
&&\hspace{+.5cm}= -\sum_n \left[ (\lambda_n+i m )
    (b'_n c_n+b_n c_n')+c.c.\right]\,.
\label{eq:actionef}\end{eqnarray}
It is left to rewrite the path integral measure as a measure of integrations 
over the coefficients $c_n,b_n$ and $c_n',b_n'$. This simple but technical 
derivation is deferred to the Appendix A. With 
the abbreviation $N_{\rm max}=1/2 \,\Tr\id$ the result is 
\begin{eqnarray}\label{eq:jacobian}
&& \prod_x d\psi_1 d\psi^*_1  d\Psi_1 d{\Psi_1}^* \nn\\
&& =  \left(
\prod_{n=1}^{N_{\rm max}} d c_{n}
 \prod_{n=1}^{N_{\rm max}} d b_{n} \prod_{n=1}^{N_{\rm max}} d c'_{n}
 \prod_{n=1}^{N_{\rm max}} d b'_{n}\right)\nn\\
&&\times \left(
\prod_{n=1}^{N_{\rm max}} d c_{n}
 \prod_{n=1}^{N_{\rm max}} d b_{n} \prod_{n=1}^{N_{\rm max}} d c'_{n}
 \prod_{n=1}^{N_{\rm max}} d b'_{n}\right)^{*}
\end{eqnarray}

Using these results we obtain
\begin{eqnarray}\label{eq:Z1}
Z= \prod_{\lambda_{n}} \left(\lambda_{n}^2 + m^{2}\right)^{2}\,.
\end{eqnarray}
Except for the zero modes and the biggest eigenvalues $\lambda_{\rm max}= 
2/a$, all eigenvalues come in pairs $\pm \lambda_n$. More explicitly, 
the related eigenfunction $\varphi^-_n$ with $H \varphi^-_n=-\lambda_n 
\varphi^-_n$ is provided by 
\begin{eqnarray}\label{eq:-eigenvalue}
\varphi^-_n= 
\frac{1}{\sqrt{1 - a^{2} \lambda_{n}^{2}/4}} \gamma_{5} (1 - a/2 D) 
\varphi_n \,.
\end{eqnarray}
Note that the above operator acting on $\varphi_n$ provides a normalised 
$\gamma_5$. It cannot be extended to all eigenfunctions because 
of the topological obstructions \cite{Jahn:2002kg}. Such a construction 
precisely fails at the largest eigenvalues $\pm \lambda_{\rm max}$, as 
can be seen also from \eq{eq:-eigenvalue}. 

Let $n_+$ and 
$n_-$ be a number of 
zero modes with positive and negative chiralities, and 
$N_+$ and $N_-$ be a number of the eigenfunctions with 
eigenvalues $\pm \lambda_{\rm max}$. Then we conclude 
\begin{equation}\label{eq:Z2}
Z=m^{2(n_++n_-)} 
\left(\0{4}{a^2} +m^2\right)^{N_++N_-} \prod_{0<\lambda_n\neq 2/a} 
(\lambda^2+m^2)^4\,, 
\end{equation} 
and for the massless limit 
\begin{eqnarray}\label{eq:Z3}
Z= \left(\0{4}{a^2}\right)^{N_++N_-} \prod_{0<\lambda_n\neq 2/a} 
\lambda_n^8\,.
\end{eqnarray} 
In other words, the partition function \eq{eq:pathmajo} can be 
 expressed in terms of Pfaffian:
\begin{eqnarray}\label{eq:Zfinal} 
Z={\rm PF}(CD)^2\,{\rm PF}(C^*D^*)^2\,, 
\end{eqnarray} 
as is mandatory for our construction. This concludes the derivation of 
a lattice action for Majorana fermions. \\[-1ex]

\noindent {\bf CP invariance in chiral gauge theories.---}
It has been widely recognised that there is an
obstruction in showing CP invariance of chiral gauge theories, see e.g
\cite{Fujikawa:2002vj,Hasenfratz:2005ch}.  Consider the standard lattice 
action of a chiral gauge theory
\begin{equation}\label{eq:CG-action1} 
S[\psi,\bar\psi,U] =
  \sum_{x,y \in \Lambda} \bar\psi(x) \Bigl(\frac{1 -
    \gamma_{5}}{2}\Bigr)
D(U)
(x-y) 
{\psi}(y) \,, 
\end{equation} 
with $\gamma_5 D(U)+D(U)\hat \gamma_5(U)=0$, and hence the chiral action in 
\eq{eq:CG-action1} has the property $\bar\psi (1-P) D\psi=\bar\psi D\hat P \psi$ with 
the chiral projection operators $P,\,{\hat P}$ defined \eq{eq:PhatP} . The 
Ginsparg-Wilson Dirac operator, $D(U)$, now depends on the link
variable for gauge field $U$, as does $\hat \gamma_5(U)$, 
\begin{equation}
  {\hat\gamma}_{5}(U) =\gamma_{5}(1-a D(U))\,.
\label{eq:hatgamma5}
\end{equation} 
We infer from \eq{eq:hatgamma5} that
also the chiral projection operator ${\hat P}$  
now depends on the link variable via $\hat\gamma_5$. 

Now we proceed to the question of a $CP$-invariant lattice action. It follows from 
the above discussion that the  no-go theorem \cite{Jahn:2002kg} prevents the construction of a chiral 
lattice action that is invariant under 
the standard $CP$-transformation in the continuum. Note however that the no-go theorem is not an obstruction for 
the construction of $CP$-invariance on the lattice as $CP$ is a discrete transformation. Still we can infer that 
a lattice generalisation of $CP$ necessarily depends on the link variable. In the following we shall consider 
non-trivial lattice generalisations of either charge conjugation or 
parity transformation. 

Here we put forward a construction where we keep the standard parity transformation and 
provide a lattice modification of charge conjugation. We emphasise again that this construction is not unique, we can 
modify both, parity and charge conjugation. This is exemplified in
Appendix B for 
modified lattice parity and standard charge conjugation. The standard parity transformations relevant in the present 
construction are
given by 
\begin{eqnarray} 
\psi(x) \to \psi^{P}(x) &=& P^{-1} \psi(x_{P})\nn\\[1ex]
\bar\psi(x) \to \bar\psi^{P}(x) &=& \bar\psi(x_{P})P\,,
\label{eq:Parity}
\end{eqnarray}
where $P$ denotes the standard parity transformation matrix, 
and $x_P=(-x_1,-x_2,-x_3,x_4)$. The properties of the link variable 
$U_\mu$ under parity transformations are summarised in 
\begin{eqnarray}\label{eq:P-link-variable}
U_{\mu}(x) \to U_{\mu}^{P}(x) =
\left\{
		\begin{array}{ll}
		 U_{i}^{\dagger}(x_{P}- a {\hat i}) & \\[1ex]
		 U_{4}(x_{P}) & {} 
		\end{array}
               \right.
\end{eqnarray}
for  $i=1,2,3$.  For the Dirac operator, we find
\begin{eqnarray} 
P D(U^{P}) P^{-1} (x,y) &=& D(U)(x_{P}, y_{P}) \,.               
\end{eqnarray}
We proceed with the observation that the natural choice for the 
charge conjugation properties of the link variable $U$ and the Dirac operator $D$ are the continuum 
properties. In the Dirac operator we have not included the chiral projections and hence both, 
$D$ and $U$, are insensitive to the topological obstructions  
related to the chiral properties of the theory. For the link variable $U$ charge conjugation reads 
\begin{eqnarray}\label{eq:CC-link-variable}
U_{\mu}(x) \to U_{\mu}^{C}(x)=\bigl(U_{\mu}^{\dagger}\bigr)^{T}(x) \,.
\end{eqnarray}
\Eq{eq:CC-link-variable} also enters the charge conjugation for the Dirac operator, which is given by  
\begin{eqnarray} 
C {D}(U^{C})  C^{-1} &=& \bigl(D(U)\bigr)^{T}\,.
\end{eqnarray}
In the case of the fermions $\psi,\bar\psi$ charge conjugation defines a mapping between 
spinors with given chirality defined in terms of the chiral projection operators $P,\,\hat P$. Therefore, we 
need to define charge conjugation
  including chiral projections as 
\begin{eqnarray} 
\bar\psi(x) \frac{1 \pm \gamma_{5}}{2} &\to& 
\left(\psi^{T} C 
\frac{1 \pm \tilde\gamma_{5}(U^{C})}{2}\right)(x)\nn\\[2ex]
\left(\frac{1 \pm \hat\gamma_{5}(U)}{2}\psi\right)(x)
 &\to& -
\frac{1 \pm \gamma_{5}}{2}C^{-1} \bar\psi^{T}(x)\,,
\label{eq:CC-new}
\end{eqnarray} 
where 
\begin{eqnarray}\label{eq:gamma5-tilde} 
\tilde\gamma_{5}(U)= (1- a~D(U))\gamma_{5}\,.
\end{eqnarray} 
As indicated before, the transformation \eq{eq:CC-new} provides a (discrete) mapping between topologically different 
spinor spaces. Note that our definition of charge conjugation \eq{eq:CC-new} reduces to
\begin{eqnarray}
\psi &\to& - C^{-1} \bar\psi^{T}\nn\\[1ex]
\bar\psi &\to& \psi^{T} C
\label{eq:CC-cont1}
\end{eqnarray}
in the continuum limit. Collecting the above results, it is straightforward to show that the action
\eq{eq:CG-action1} is CP invariant. 
Note that the functional measure can be constructed as put forward in \cite{Luscher:1998du}, using the
chiral projection operators $P,\,{\hat P}$. It 
is invariant under the CP transformation with \eq{eq:CC-new} and \eq{eq:Parity}.

Our lattice extension of the charge conjugation 
also applies to the symplectic Majorana fermions: 
\begin{eqnarray} 
\psi_{a}^{T}(x)C \frac{1 \pm \gamma_{5}}{2} 
& \to &\varepsilon_{ab}\left(\Psi_{b}^{T}C
\frac{1 \pm \tilde\gamma_{5}(U^{C})}{2}\right)(x)
\label{eq:CC-MW}\nn\\[2ex]
\left(\frac{1 \pm \hat\gamma_{5}(U)}{2}\Psi_{a}\right)(x)
& \to &\varepsilon_{ab}
\frac{1 \pm \gamma_{5}}{2} \psi_{b}(x)\,,
\end{eqnarray}  
where $a,b =1 ~{\rm or}~ 2$. 
Parity transformation reads
\begin{eqnarray}
\Psi_{1}(x) \to \Psi^{P}_{1}(x) &=& \eta ~P^{-1} \Psi_{1}(x_{P})\nn\\[1ex]
\Psi_{2}(x) \to \Psi^{P}_{2}(x) &=& - \eta ~P^{-1} \Psi_{2}(x_{P})
\label{eq.P-MW}\\[2ex]
\psi_{1}^{T}(x)C \to \left(\psi_{1}^{T}(x)C\right)^{P}(x) &=& \eta ~\psi_{1}^{T}(x_{P})C P\nn\\[1ex]
\psi_{2}^{T}(x)C \to \left(\psi_{2}^{T}(x)C\right)^{P}(x) &=& - \eta
 ~\psi_{2}^{T}(x_{P})C P \nn\,.
\end{eqnarray}
where $\eta=\pm 1$. Then, we can show that a chiral gauge theory described 
by the Majorana-Weyl action 
\begin{equation}
S_{MW}[\psi,\Psi] =\sum_{x,y \in \Lambda} {\psi}_{2}^{T}(x)C 
\Bigl(\frac{1 {-} \gamma_{5}}{2}\Bigr) {D}(U)(x-y) {\Psi}_{1}(y)\,.
\label{eq:MW-action}
\end{equation}
is CP invariant. 
\\[-1ex]

\noindent {\bf Conclusion.---} We have shown that the
construction of a theory with chirally coupled Majorana fermions on
the lattice has to deal with the usual topological obstructions
well-known from the construction of chiral fermions, even though the
total chirality is even.  The obstruction is related to the use of
chiral projection operators in the Yukawa term. This problem is
resolved by doubling the degrees of freedom, and the Pfaffian nature
of the path integral is proven.  

We have also shown that the 
difficulty of constructing CP invariant chiral gauge theories arises
from the same topological obstruction. In order to derive a
CP invariant action for a chiral gauge theory, we introduce a lattice
generalisation of the CP transformation. This was done by modifying either 
lattice parity or charge conjugation or both. The construction also applies to 
Majorana-Weyl fermions. It remains to be seen how amiable these
constructions are towards numerical implementation, see
  e.g.\
  \cite{Montvay:2001aj,Catterall:2001fr,Beccaria:2004ds,Kastner:2008zc,%
Giedt:2008cd,Kanamori:2008bk, Endres:2009}. To that end
one also should further explore the generality of the approach
presented here in order to have maximal flexibility within the
numerical implementation.  Another interesting direction is the
extension of the construction presented here to supersymmetric
theories.  In the continuum theory, symplectic Majorana fermions can
be used to discuss extended supersymmetric theories. It is therefore
challenging to construct such theories on the lattice. \\[-2ex]

\noindent {\bf Acknowledgements.---} YI would like to thank 
the Institute of Theoretical Physics in
Heidelberg for hospitality. We are grateful to F.~Bruckmann and 
H.~Suzuki for various useful discussions.\\[-6ex]

\renewcommand{\thesection}{}
\renewcommand{\thesubsection}{\Alph{subsection}}
\renewcommand{\theequation}{A.\arabic{equation}}

\subsection{A. Path integral measure}\label{app:derofmeasure}
\setcounter{equation}{0}
\vspace{-.3cm}

Here we rewrite the path integral measure over $\psi,\Psi$ in terms of 
the expansion coefficients $c_n,b_n$ and $c_n',b_n'$. We start with  
\begin{eqnarray}\label{eq:measure}
\prod_{x} d \psi_{1} = \det V^{-1} \prod_{n=1}^{N_{\rm max}} 
dc_{n} \prod_{n=1}^{N_{\rm max}} db_{n}\,,
\end{eqnarray}
where $N_{\rm max}=1/2\,\Tr\id$, and the matrix $V$ is defined by
\begin{eqnarray}\label{eq:matrixV}
V_{x,n} = \left(\varphi_{n}(x), \gamma_5 C^{-1} \varphi_n^*(x)\right)\,. 
\end{eqnarray}
The matrix $V$ has the properties 
\begin{equation}\label{eq:Vrelation1}
\sum_{x,l}V^{T}_{m,x}B^{-1}V_{x,l}
\left(
\begin{array}{cc}
0 & -\delta_{l,n}\\
\delta_{l,n} & 0
\end{array}
\right) 
= 
\left(
\begin{array}{cc}
\delta_{m,n} & 0\\
0 & \delta_{m,n}
\end{array}
\right)\,,
\end{equation}
and 
\begin{eqnarray}\label{eq:Vrelation2}
(\det V)^{-2} = (-)^{N_{\rm max}^{2} +N_{\rm max}^{2} } 
\det (\gamma_{5}C^{-1})\,. 
\end{eqnarray}
With help of these relations we conclude that \eq{eq:jacobian} holds.

\renewcommand{\theequation}{B.\arabic{equation}}

\subsection{B. Modified lattice parity}\label{app:modpar}
\setcounter{equation}{0}
Let us discuss a lattice generalisation of parity transformation,
realising charge conjugation in the standard way. For the action
\eq{eq:CG-action1}, we adopt the charge conjugation
\eq{eq:CC-cont1}. 
The parity transformation is required to give a
mapping between topologically different spinor states with opposite
chiralities. Therefore, we need to introduce chiral
projection operators explicitly for a lattice generalisation of the parity
transformations, 
\begin{eqnarray}\nn
\frac{1\pm \gamma_{5}}{2}\psi(x) &\to& 
P^{-1} \left(\frac{1\mp \hat\gamma_{5}(U)}{2}\psi\right)(x_{P}) \\[2ex]
\left(\bar\psi \frac{1\pm
				     \tilde\gamma_{5}(U)}{2}\right)(x) 
&\to &\bar\psi(x_{P})  \frac{1\mp \gamma_{5}}{2} P 
\label{eq:P-new1}
\,.
\end{eqnarray}
The action \eq{eq:CG-action1} is shown to be invariant under combined 
operations \eq{eq:CC-cont1} and \eq{eq:P-new1}.

Let us consider the Majorana-Weyl action \eq{eq:MW-action}. 
The parity transformations should flip the 
chirality which is defined with different projection operators for
$\psi$ and $\Psi$.  It reads for the fermions $\psi$, 
\begin{eqnarray}
\frac{1 \pm \gamma_{5}}{2} \psi_{1}(x) 
&\to&  
    \eta P^{-1} \left(\frac{1 \mp 
      \hat\gamma_{5}(U)}{2} \Psi_{1}\right)(x_{P}) \hspace{1cm}\ \label{P-1} \\[2ex]
\frac{1 \pm \gamma_{5}}{2} \psi_{2}(x)  &\to& 
    -\eta P^{-1} \left(\frac{1 \mp \hat\gamma_{5}(U)}{2} \Psi_{2}\right)(x_{P})\nn
\,.
\end{eqnarray}
For the fermion $\Psi$ we define 
\begin{eqnarray}
\left(\frac{1 \pm \hat\gamma_{5}(U)}{2}
\Psi_{1}\right)(x) &\to& 
\eta P^{-1} \frac{1 \mp \gamma_{5}}{2} 
\psi_{1}(x_{P})\nn\\[2ex]
\left(\frac{1 \pm \hat\gamma_{5}(U)}{2}
\Psi_{2}\right)(x) &\to& 
-\eta  P^{-1}  \frac{1 \mp \gamma_{5}}{2} \psi_{2}(x_{P})\nn\\
\label{P-2}\,,
\end{eqnarray}
Collecting the above results and using  \eq{eq:CC-cont1}, it is straightforward to show that 
Majorana-Weyl action \eq{eq:MW-action} is CP invariant.

\end{document}